\documentclass[aps,prc,preprint,showpacs,superscriptaddress]{revtex4}
\usepackage{epsfig}
\usepackage{amsmath}
\usepackage[hyperref]{hyperref}

\begin{document}

\title{Electromagnetic and strong contributions to dAu soft coherent
inelastic diffraction at RHIC}

\author{V. Guzey}
\email{vguzey@jlab.org}
\affiliation{Theory Center, Jefferson Lab, Newport News, VA 23606, USA}

\author{M. Strikman}
\email{strikman@phys.psu.edu}
\affiliation{Department of Physics, Pennsylvania State University, University
Park, PA 16802, USA}

\preprint{JLAB-THY-08-815}

\begin{abstract}

We estimate electromagnetic (ultra-peripheral) and strong contributions to dAu soft coherent
inelastic diffraction at RHIC, $dAu \to X Au$.
We show that the electromagnetic contribution is the dominant one and that
the corresponding cross section is sizable, $\sigma^{dAu \to XAu}_{\rm e.m.}=214$ mb, which
constitutes 10\% of the total dAu inelastic cross section.

\end{abstract}

\maketitle


Deuteron -- Gold (dAu) collisions constitute an essential part of the present physics 
program of the Relativistic Heavy Ion Collider (RHIC). 
Ultra-peripheral collisions (UPC) of ions of deuterium and Gold (or any other ions) 
are processes, where the colliding nuclei are separated by 
the transverse distance (impact parameter) larger than the sum of the deuteron and Gold
radii. In this case, the heavier ion of Au acts as a source of quasi-real photons 
of a very high energy (the photon flux produced by deuterium is negligibly small 
compared to that by Au)
and, hence, one effectively studies the interaction of photons with deuterons,
$dAu \to d\gamma Au \to X Au$~\cite{Baltz:2007kq,Baltz:2007hw}.

At RHIC, dAu ultra-peripheral collisions were studied in the following channels:
$dAu \to d Au \rho^0$ and $dAu \to np Au \rho^0$~\cite{Timoshenko:2007ti},
and $dAu \to np Au$~\cite{White:2005rt}. The cross section of the latter reaction
 was normalized to the theoretical prediction,
$\sigma^{dAu \to np Au}=1.38$~b~\cite{Klein:2003bz}.
This value is large and should be compared to the measured 
total dAu inelastic cross section, $\sigma^{dAu \to X}=2.26$~b~\cite{Baltz:2007kq}.
Since  $\sigma^{dAu \to np Au}$ is so large, the $dAu \to np Au$ yield was used to
determine the absolute normalization of the RHIC dAu data.

In this paper, we consider yet another channel of dAu UPC, namely,
deuteron inelastic diffraction,
$dAu \to d\gamma Au \to X Au$, where $X$ denotes products of the deuteron inelastic
dissociation. 
We find that the corresponding cross section is sizable,
 $\sigma^{dAu \to XAu}_{\rm e.m.}=214$ mb, which
constitutes 10\% of the total dAu inelastic cross section.

Although the discussed cross section is 
only  $\sim 10$\% of the total inelastic cross section, it may contribute a larger fraction of the signal in certain cases. For example, if one considers the leading pion
or nucleon production at small transverse momenta $p_t$, one observes that the multiplicity of such processes is about the same in $pN$ and $\gamma N$ interactions. 
At the same time, the $A$-dependence of the forward hadron multiplicity 
for 
$x_{F} > 0.4$ ($x_F$ is the fraction of the projectile's momentum carried by the leading hadron)
is  $\approx A^{-1/3}$. 
Hence the $\gamma N$ and $pN$ interactions can give comparable contributions in
the considered case.

We also estimate the strong contribution to $dAu  \to X Au$
soft coherent inelastic diffraction and find that the corresponding
cross section is rather small, $\sigma^{dAu \to XAu}_{\rm dd}=22$ mb.


The cross section of deuteron inelastic diffraction in dAu ultra-peripheral scattering
reads, see e.g.~\cite{Baltz:2007kq,Baltz:2007hw},
\begin{equation}
\sigma^{dAu \to XAu}_{\rm e.m.}(s)=2\,\int^{\omega_{\rm max}}_{\omega_{\rm min}} d \omega\,
\frac{dN_{\gamma}(\omega)}{d \omega}\, \sigma_{\rm tot}^{\gamma p}(\omega)
 \,,
\label{eq:cs}
\end{equation}
where $dN_{\gamma}(\omega)/d \omega$ is the flux of equivalent photons emitted by Au;
 $\sigma_{\rm tot}^{\gamma p}(\omega)$ is the total real photon-nucleon cross section;
$s$ is the total invariant energy squared per nucleon ($\sqrt{s}=200$ GeV at the present
RHIC energy);
$\omega$ is the photon energy in the deuteron rest frame; $\omega_{\rm max}$ and $\omega_{\rm min}$ are the 
maximal and minimal values of $\omega$ (see below).
 The factor of two is the reflection that
the total photon-deuteron cross section is twice the total photon-proton
cross section. Note that Eq.~(\ref{eq:cs}) is implicitly 
invariant with respect to boosts along the collision axis.
We will work in the deuteron rest frame.

In Eq.~(\ref{eq:cs}), the flux of equivalent photons emitted by the fast nucleus
of Au is~\cite{Baltz:2007kq,Baltz:2007hw}
\begin{eqnarray}
\frac{dN_{\gamma}(\omega)}{d \omega}&=&\frac{Z^2 \alpha \omega}{\pi^2 \gamma^2} \int_{|b| > R_A+R_d} d^2b \left[K_1^2 \left(\frac{\omega |b|}{\gamma}\right)+\frac{1}{\gamma^2}K_0^2 \left(\frac{\omega |b|}{\gamma}\right) \right]  \nonumber\\
&=& \frac{2\,Z^2 \alpha}{\pi \omega}\left[x K_0(x)K_1(x)+\frac{x^2}{2}
\left(K_0^2(x)-K_1^2(x) \right) \right]
\,,
\label{eq:flux}
\end{eqnarray}
where $Z$ is the electric charge ($Z=79$ for Au); 
$\alpha \approx 1/137$ is
the fine-structure constant;
 $\gamma$ is the Lorentz factor of the fast moving Au; 
$b$ is the distance between the centers of Au and the deuteron
in the transverse plane (impact parameter), which should be larger than
the sum of the 
corresponding Au and deuteron radii, $R_A$ and $R_d$;
$K_0$ and $K_1$ are modified Bessel functions; $x=(R_A+R_d) \omega/ \gamma$.
Note also that we omitted the negligibly small contribution of 
the $K_0^2/\gamma^2$ term, when going from the first to the second line
 of Eq.~(\ref{eq:flux}).

The maximal energy of the exchanged photon, 
$\omega_{\rm max}$, is determined by the Lorentz-contracted
nuclear size,  $\omega_{\rm max}=\gamma/R_A$.
 The minimal photon energy,
$\omega_{\rm min}$, is termined by the threshold of the inelastic
$\gamma+p \to \Delta(1232)$ reaction. In the proton rest frame, 
$\omega_{\rm min}=0.3$ GeV.

For the evaluation of $\sigma^{dAu \to XAu}_{\rm e.m.}$ using Eq.~(\ref{eq:cs}),
we used the following input.
At the current RHIC energy of $\sqrt{s}=200$ GeV per nucleon,
the Lorentz factor of Au in the deuteron rest frame is 
$\gamma=2.3 \times 10^5$. The Au ($A=197$) effective radius was parameterized in 
a simple form, $R_A=1.145\,A^{1/3} \approx 6.7$ fm.
Therefore, the maximal photon energy is $\omega_{\rm max}=670$ GeV.
 For the deuteron radius, we
used $R_d=1.88$ fm, which was obtained using the deuteron wave function
with the Paris nucleon-nucleon potential~\cite{Lacombe:1981eg}.
We checked that the result of Eq.~(\ref{eq:cs_num}) does not change, 
if we vary $R_d$ by 1 fm in order to mimic the effect 
of fluctuations of the size of the proton-neutron system in the deuteron.
For the real photon-proton cross section, we used
the ALLM parameterization~\cite{Abramowicz:1991xz}.

\begin{figure}[h]
\begin{center}
\epsfig{file=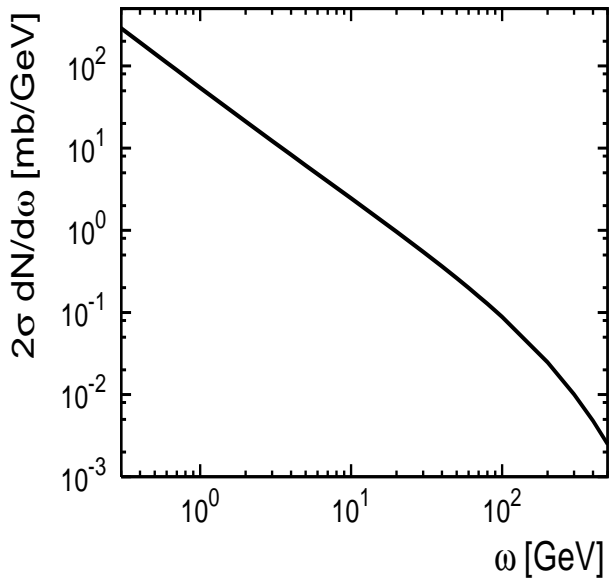,width=9cm,height=8cm} 
\caption{The integrand of Eq.~(\ref{eq:cs}),
$2 \sigma_{\rm tot}^{\gamma p}(\omega)dN_{\gamma}(\omega)/d \omega$,
as a function of the photon energy $\omega$ in the deuteron rest frame.}
\label{fig:Integrand}
\end{center}
\end{figure}

Figure~\ref{fig:Integrand} presents the integrand of Eq.~(\ref{eq:cs}),
$2 \sigma_{\rm tot}^{\gamma p}(\omega)dN_{\gamma}(\omega)/d \omega$,
as a function of the photon energy $\omega$ in the deuteron rest frame.

A direct evaluation of Eq.~(\ref{eq:cs}) using the input described above gives
\begin{equation}
\sigma^{dAu \to XAu}_{\rm e.m.}=214 \ {\rm mb} 
 \,.
\label{eq:cs_num}
\end{equation}
This value is quite sizable: it constitutes 10\% of the total dAu 
inelastic cross section, $\sigma^{dAu \to X}=2.26$ b~\cite{Baltz:2007kq}.

The application of Eq.~(\ref{eq:cs}) to inelastic diffraction in deuteron-Lead UPC at the LHC energy ($\sqrt{s}=6.2$ TeV) gives
$\sigma^{dPb \to XPb}_{\rm e.m.}=828$ mb.


The value of $\sigma^{dAu \to XAu}_{\rm e.m.}$
can be compared to the cross section of $dAu \to XAu$ soft coherent inelastic diffraction,
$\sigma^{dAu \to XAu}_{\rm dd}$, which is due to the strong interaction.
This cross section can be calculated based on the concept of
 cross section fluctuations~\cite{FP56,Good:1960ba}. 
Within this formalism, an analysis of various phenomena involving 
proton-proton and proton-deuteron scattering allowed to determine 
the distribution over the strength of the interaction~\cite{Blaettel:1993ah}, 
which lead to the explanation of cross sections
of inelastic diffraction of protons and pions off 
nuclei at fixed-target energies~\cite{Frankfurt:1993qi,Strikman:1995jf}. 
 For a summary and extension to collider energies, 
see~\cite{Frankfurt:2000tya,Guzey:2005tk}.
Below we outline the derivation of $\sigma^{dAu \to XAu}_{\rm dd}$.

In the standard Glauber method~\cite{Glauber:1955qq,Franco:1969sn},
 inelastic diffraction is absent and the
only two allowed diffractive final states in coherent dAu scattering are elastic, 
$d Au \to d Au$, and deuteron dissociation, $d Au \to pn Au$. Using the 
completeness of these two diffractive states, the sum of the corresponding cross
sections can be expressed as
\begin{eqnarray}
&&\sum_{X=d,pn}\sigma^{dAu \to X Au}= \int d^2b\, d^2 r_t \,|\psi_D(r_t)|^2 \nonumber\\
& \times & 
\left|\left(1-\exp\left[-\frac{\sigma^{pN}_{\rm tot}}{2}\,T\left(b+\frac{r_t}{2}\right)
-\frac{\sigma^{nN}_{\rm tot}}{2}\,T\left(b-\frac{r_t}{2}\right)
+\sigma_{\rm el}^{NN} e^{-r_t^2/(4B_{\rm el})} T(b) \right]\right)\right|^2 \,.
\label{eq:glauber}
\end{eqnarray}
In this equation, $b$ is the transverse distance between the centers of Au and d;
$r_t$ is the transverse distance between the proton and neutron in the deuteron;
$\psi_D(r_t)$ is the deuteron wave function; $\sigma^{pN}_{\rm tot}$ and
$\sigma^{nN}_{\rm tot}$ are respectively the proton-nucleon and neutron-nucleon total scattering cross sections (for the purpose of the following discussion, it is 
convenient to distinguish between the two cross sections);  
$\sigma_{\rm el}^{NN}$ is the elastic nucleon-nucleon (NN) cross section;
$B_{\rm el}$ is the slope of the NN elastic amplitude; $T(b)$ is the so-called 
optical density of the nucleus of Au, $T(b)=\int dz \rho_A(r)$, where 
$\rho_A(r)$ is the nuclear density~\cite{De Jager:1987qc}.
The real part of the NN scattering amplitude is neglected in 
Eq.~(\ref{eq:glauber}).

In the following analysis, we neglect the contribution of the last term
in the exponent in Eq.~(\ref{eq:glauber}), which is suppressed by the smallness
of the elastic cross section, $\sigma_{\rm el}^{NN}/\sigma_{\rm tot}^{NN} \approx 1/5$
at $\sqrt{s}=200$ GeV, and by the smallness of the 
$e^{-r_t^2/(4B_{\rm el})}$ factor, 
$e^{-r_t^2/(4B_{\rm el})} \approx 0.15$ at $r_t=2$ fm and $B_{\rm el}=13$ GeV$^{-2}$.

In order to take into account inelastic diffraction of deuterons,  
the standard Glauber method can be complemented by the formalism of 
cross section fluctuations. In the combined approach, the $dAu \to X Au$
coherent inelastic diffractive (diffraction dissociation) cross section 
reads
\begin{eqnarray}
&&\sigma^{dAu \to X Au}_{\rm dd}= \int d^2b\, d^2 r_t \,|\psi_D(r_t)|^2 \nonumber\\
& \times & 
\Bigg\{\int d\sigma_p P(\sigma_p) d\sigma_n P(\sigma_n)\left| \left(1-\exp\left[-\frac{\sigma_p}{2}\,T\left(b+\frac{r_t}{2}\right)
-\frac{\sigma_n}{2}\,T\left(b-\frac{r_t}{2}\right) \right]\right)\right|^2 
 \nonumber\\
& - & \left|\int d\sigma_p P(\sigma_p) d\sigma_n P(\sigma_n) \left(1-\exp\left[-\frac{\sigma_p}{2}\,T\left(b+\frac{r_t}{2}\right)
-\frac{\sigma_n}{2}\,T\left(b-\frac{r_t}{2}\right) \right]\right)\right|^2\Bigg\}
\,,
\label{eq:glauber_plus_csf}
\end{eqnarray}
where the proton and neutron of the deuteron interact with the target nucleons with
the total cross sections $\sigma_p$ and $\sigma_n$, whose probability distributions
are given by $P(\sigma_p)$ and  $P(\sigma_n)$, respectively.
The distribution $P(\sigma)$ is peaked around the total NN cross section
with a small dispersion. This allows one to expand the exponents 
in Eq.~(\ref{eq:glauber_plus_csf}) around $\sigma_{\rm tot}^{NN}$~\cite{Frankfurt:1993qi}
and to obtain
\begin{eqnarray}
\sigma^{dAu \to X Au}_{\rm dd}&=&\frac{(\sigma_{\rm tot}^{NN})^2\, \omega_{\sigma}}{4}\,
2
\int d^2b\, d^2 r_t \,|\psi_D(r_t)|^2 \left[T\left(b+\frac{r_t}{2}\right)\right]^2
\nonumber\\
&\times&\exp\left[-\sigma_{\rm tot}^{NN}\left(T\left(b+\frac{r_t}{2}\right)+T\left(b-\frac{r_t}{2}\right)\right)\right] \,,
\label{eq:glauber_plus_csf_approx}
\end{eqnarray}
where $\omega_{\sigma}$ is a parameter describing the dispersion of $P(\sigma)$ around
its maximum and is proportional to $\sigma^{NN \to XN}_{\rm dd}$,
the cross section of inelastic diffraction in NN scattering.
At $\sqrt{s}=200$ GeV, $\omega_{\sigma}$ is maximal, 
$\omega_{\sigma}=0.3$~\cite{Guzey:2005tk}.
The factor of two in front of the integral in Eq.~(\ref{eq:glauber_plus_csf_approx})
is a sum of the proton and neutron contributions.
Note that we do not distinguish the proton-nucleon and neutron-nucleon 
cross sections in Eq.~(\ref{eq:glauber_plus_csf_approx}).

\begin{figure}[h]
\begin{center}
\epsfig{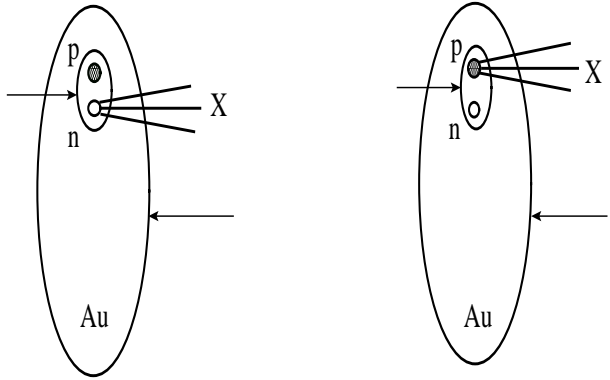} 
\caption{The schematic representation of $dAu \to XAu$ coherent inelastic
diffraction as
a sum of two contributions, when either the neutron (left picture) or the proton (right picture) diffracts inelastically.}
\label{fig:dAu_dd_graph}
\end{center}
\end{figure}

The physical interpretation of Eq.~(\ref{eq:glauber_plus_csf_approx}) 
is the following. 
Inelastic diffraction of deuterons receives independent
and equal contributions from inelastic diffraction of the proton and neutron
of the deuteron. The contribution, when both the proton and neutron
diffract, is small and has been neglected. 
Let us consider the contribution due to the 
proton diffraction. At high energies, the internal motion of nucleons in the
deuteron and in the nucleus of Au is Lorentz dilated and, hence, the nucleons can be considered ''frozen''
in the their transverse positions. The proton of the deuteron at the position $\vec{b}+\vec{r_t}/2$ undergoes inelastic diffraction with the probability
proportional to $[T\left(b+r_t/2)\right]^2\,\exp(-\sigma_{\rm tot}^{NN}T\left(b+r_t/2\right))$. At the same time, the neutron at the position $\vec{b}-\vec{r_t}/2$
does not take part in the 
interaction. The probability for the neutron to not interact is 
$\exp(-\sigma_{\rm tot}^{NN}\,T\left(b-r_t/2\right))$. An equal contribution 
to the deuteron inelastic diffraction comes from the situation, when the 
proton and neutron switch roles. The two contributions are schematically presented
in Fig.~\ref{fig:dAu_dd_graph}.

A direct evaluation of Eq.~(\ref{eq:glauber_plus_csf_approx}) gives
\begin{equation}
\sigma^{dAu \to X Au}_{\rm dd}=22 \ {\rm mb} \,.
\end{equation}
In our calculation, we used the Paris deuteron wave function~\cite{Lacombe:1981eg} 
and the parameterization of $\sigma^{NN}_{\rm tot}$ due to Donnachie and Landshoff~\cite{Donnachie:1992ny}.

We would like to note that various channels of dAu scattering in the RHIC kinematics
using the Glauber method combined with the dipole formalism (which leads to 
cross section fluctuations) were considered in~\cite{Kopeliovich:2003tz}. 
However, soft coherent inelastic diffraction
or ultra-peripheral collisions were not addressed.

In summary, we estimated deuteron inelastic diffraction in ultra-peripheral 
$dAu \to d\gamma Au \to X Au$ scattering, when the ion of Au serves as a source
of high-energy photons.
We found that the corresponding cross section is sizable,
 $\sigma^{dAu \to XAu}_{\rm e.m.}=214$ mb, which
constitutes 10\% of the total dAu inelastic cross section.
The same reaction can also proceed through the strong interactions.
We derived an approximate expression for the cross section of $dAu \to X Au$
soft coherent inelastic diffraction and estimated it to be rather small,
$\sigma^{dAu \to X Au}_{\rm dd}=22$ mb.

We also estimated the cross section of deuteron-Lead inelastic diffraction in 
ultra-peripheral $dPb \to d\gamma Pb \to X Pb$ scattering at the LHC energies ($\sqrt{s}=6.2$ TeV) and found
$\sigma^{dPb \to XPb}_{\rm e.m.}=828$ mb. The corresponding cross section of
$dPb \to X Pb$ soft coherent inelastic diffraction due to the strong
interactions is negligibly small,
$\sigma^{dAu \to X Au}_{\rm dd}=7$ mb,
 due to the vanishingly small $NN \to XN$ inelastic diffraction.

We want to thank Sebastian White for useful discussions and Boris Kopeliovich
for explanations of the results of Ref.~\cite{Kopeliovich:2003tz}.
The work of M.S. is supported by DOE.\\
Notice: Authored by Jefferson Science Associates, LLC under U.S. DOE Contract No. DE-AC05-06OR23177. The U.S. Government retains a non-exclusive, paid-up, irrevocable, world-wide license to publish or reproduce this manuscript for U.S. Government purposes.

\end{document}